\documentclass[aps,prb,twocolumn,longbibliography,preprintnumbers,amsmath,amssymb,superscriptaddress]{revtex4-1}

\usepackage{graphicx}
\usepackage[T1]{fontenc}
\usepackage[colorlinks,bookmarks=false,citecolor=blue,linkcolor=red,urlcolor=blue]{hyperref}
\usepackage[dvipsnames]{xcolor}
\usepackage{times}

\newcommand{\be}{\begin{eqnarray}}
\newcommand{\ee}{\end{eqnarray}}

\newcommand{\nn}{\nonumber } 
\newcommand{\Eqref}[1]{Eq.~\eqref{#1}}

\begin{document}

\author{Daniel D. Scherer}\email{daniel.scherer@nbi.ku.dk}
\affiliation{Niels Bohr Institute, University of Copenhagen, DK-2100 Copenhagen, Denmark}

\author{Ilya Eremin}
\affiliation{Institut fur Theoretische Physik III, Ruhr-Universitat Bochum, D-44801 Bochum, Germany}

\author{Brian M. Andersen}
\affiliation{Niels Bohr Institute, University of Copenhagen, DK-2100 Copenhagen, Denmark}

\title{Collective magnetic excitations of $C_{4}$ symmetric magnetic states in iron-based superconductors}

\begin{abstract}
We study the collective magnetic excitations of the recently discovered $C_{4}$ symmetric spin-density wave states of iron-based superconductors with particular emphasis on their orbital character based on an itinerant multiorbital approach. This is important since the $C_{4}$ symmetric spin-density wave states exist only at moderate interaction strengths where damping effects from a coupling to the continuum of particle-hole excitations strongly modifies the shape of the excitation spectra compared to predictions based on a local moment picture. We uncover a distinct orbital polarization inherent to magnetic excitations in $C_{4}$ symmetric states, which provide a route to identify the different commensurate magnetic states appearing in the continuously updated phase diagram of the iron-pnictide family.
\end{abstract}

\maketitle

\textit{Introduction.} 
In the iron-based superconductors (FeSC), superconductivity appears in
close proximity to a magnetic instability~\cite{Chubukov2015}. Therefore, much of the research of these compounds is devoted
to understanding the magnetic properties of these systems.
Experimentally, the magnetic order of most iron pnictides has orthorhombic ($C_{2}$) symmetry and corresponds
to stripes of parallel spins modulated either along the $\hat{\mathbf{x}}$ or along
the $\hat{\mathbf{y}}$ direction with ordering wavevector $\mathbf{Q}_{1}=\left(\pi,0 \right)$ or $\mathbf{Q}_{2}=\left(0,\pi\right)$, respectively. Correspondingly, this magnetic stripe (MS) state also breaks $Z_2$ Ising-like symmetry in addition to the continuous O(3) spin-rotational symmetry broken below the magnetic transition temperature, $T_N$. The $Z_2$ (or, equivalently,
$C_2$) symmetry breaking occurs in some cases at temperatures, $T_s$,  which are higher than $T_N$ and necessitates a structural transition from tetragonal to orthorhombic symmetry. This allows for an intermediate
phase, sometimes called Ising nematic, with only broken $Z_2$ symmetry without magnetic long-range order~\cite{Fang2008,Xu2008,Fernandes2014}. On the theory side, this state has been described by a variety of approaches ranging 
from purely localized Heisenberg spins~\cite{Fang2008,Xu2008,Kruger2009,Abrahams2011,Kamiya2011}
to itinerant nesting-based scenarios~\cite{Lorenzana2008,Cvetkovic2009,Brydon2009a,Eremin2010,Fernandes2012} and 
to hybrid models mixing local moments and itinerant carriers~\cite{Dai2012,Yin2010,Lv2010,Ducatman2014,Johannes2009}.
Despite the success of these approaches in describing many magnetic properties of the iron pnictides, the fundamental
question as to the relevance of the related spin, charge or orbital fluctuations remains open. 

Recently, the tetragonal magnetic phase, preserving the $C_4$ symmetry of the lattice has been observed in the hole-doped iron pnictides~\cite{Kim10,Hassinger2012,Avci2014,Boehmer2015,Allred2015,Allred2016}, suggesting that this phase is a generic feature in the phase diagram of the FeSC. Such a state can be understood as the superposition of two spin-density waves ${\bf M}({\bf r}) = {\bf M}_1 \mathrm{e}^{\mathrm{i}{\bf Q_1}\cdot{\bf r}} + {\bf M}_2 \mathrm{e}^{\mathrm{i}{\bf Q_2}\cdot{\bf r}}$ of the original striped antiferromagnetic state. The existence of this so-called double-Q magnetic state as an additional ground state for the FeSC has also been proposed by various theoretical approaches ~\cite{Lorenzana2008,Eremin2010,Brydon2011,Giovanetti2011,Kang2011,Gastiasoro2014}. In particular, most of these studies pointed out two possible double-Q ground states. One of them is the so-called spin charge density wave order (SCO) that arises from aligning ${\bf M}_1$ and ${\bf M}_2$ either parallel or antiparallel with resulting nonuniform magnetization with vanishing average moment at the even lattice sites and staggered-antiferromagnetic order at the odd lattice sites. The other possible state is when ${\bf M}_1$ is orthogonal to ${\bf M}_2$ with a non-collinear magnetization, referred to as the othomagnetic (OM) phase.

Recent M\"ossbauer spectroscopy points in favor of a non-uniform magnetization in this ground state, which appears if ${\bf M}_1$ is either parallel or antiparallel to ${\bf M}_2$~\cite{Allred2016}. Such a state appears naturally within an itinerant description of the magnetism in FeSC, where the sizes of the magnetic moments on one sublattice can be changed by the costs of the other with additional change density wave, which develops on the non-magnetic sublattice~\cite{Kang2011,Gastiasoro2015}. Furthermore, recent analysis of the effective low-energy model, derived from the ten-orbital tight-binding model including the spin-orbit coupling, has shown the orientation of the magnetization of this phase along $\hat{\mathbf{z}}$-direction~\cite{Christensen2015} in agreement with polarized neutron scattering~\cite{Wasser15}.  

One of the interesting questions with respect to the double-Q states is the peculiarities of the spin dynamics of these phases. Previously the spin dynamics of the $C_2$ symmetric striped antiferromagnetic state was  described either within an itinerant RPA-type description~\cite{Knolle2010,Brydon2009b,Kaneshita2010,Kovacic2015}, dynamical mean-field theory approach~\cite{Yin2014} or localized spin models~\cite{Fang2008,Dai2012,Xu2008}. More recently the spin dynamics of the $C_4$ OM phase was also discussed theoretically using spin-wave theory of a Heisenberg-like Hamiltonian~\cite{Wang2015}. The itinerant nature of the $C_4$ symmetric phases, however, calls for the corresponding description of its magnetic excitations. Here, we extend previous multiorbital RPA calculations~\cite{Knolle2010,Kaneshita2010,Kovacic2015} to compute the spin excitations of the $C_4$ symmetric OM and SCO ordered phases.

\textit{Multiorbital model and magnetic phase diagram.}
The itinerant electron system of the parent FeSC is described by a multiorbital Hubbard Hamliltonian $ H = H_{0} + H_{\mathrm{int}} $, which consists of the non-interacting hopping Hamiltonian within the $3d$-orbital manifold,
\be
\label{eq:hopping}
H_{0} = \sum_{\sigma}\sum_{i,j}\sum_{\mu,\nu} c_{i \mu \sigma}^{\dagger}\left( t_{ij}^{\mu\nu}  - \mu_{0} \delta_{ij}\delta_{\mu\nu} \right)c_{j \nu \sigma}, 
\ee
and a Hubbard-Hund interaction term
\be
\label{eq:interaction}
H_{\mathrm{int}} & = &  
U \sum_{i,\mu} n_{i \mu \uparrow} n_{i \mu \downarrow} + 
\left(U^{\prime} - \frac{J}{2}\right) \sum_{i,\mu < \nu, \sigma,\sigma^{\prime}} n_{i \mu \sigma} n_{i \nu \sigma^{\prime}} \nn \\
& & \hspace{-2.5em} 
- 2 J \sum_{i, \mu < \nu}{\bf S}_{i\mu}\cdot{\bf S}_{i\nu}  + 
+J^{\prime} \sum_{i, \mu < \nu,\sigma} c_{i\mu\sigma}^{\dagger}c_{i\mu\bar{\sigma}}^{\dagger}c_{i\nu\bar{\sigma}}c_{i\nu\sigma}.
\ee
Here, the indices $\mu,\nu \in \{d_{xz}, d_{yz}, d_{x^2-y^2}, d_{xy}, d_{3z^{2}-r^{2}}\}$ specify the $3d$-Fe orbitals and $i,j$ run over the sites of the square lattice. The filling is fixed by the chemical potential $\mu_{0}$, and the onsite interaction is parametrized by an intraorbital Hubbard-$U$, an interorbital coupling $U^{\prime}$, Hund's coupling $J$ and pair hopping $J^{\prime}$. We will restrict ourselves to spin-rotational symmetric interaction parameters, which are realized for $U^{\prime} = U - 2J$, $J = J^{\prime}$. We further put $J = U/4$ in the rest of this work. The fermionic operators $ c_{i \mu \sigma}^{\dagger}$, $c_{i \mu \sigma}$ create and destroy, respectively, an electron at site $i$ in orbital $\mu$ with spin polarization $\sigma$. Accordingly, we define the operators for local charge and spin as $n_{i\mu} = n_{i\mu\uparrow} + n_{i\mu\downarrow}$ with $n_{i\mu\sigma} = c_{i\mu\sigma}^{\dagger} c_{i\mu\sigma}$ and ${\bf S}_{i\mu} = 1/2\sum_{\sigma\sigma^{\prime}} c_{i\mu\sigma}^{\dagger} {\boldsymbol \sigma}_{\sigma\sigma^{\prime}}c_{i\mu\sigma^{\prime}}$, respectively. 

\begin{figure}[t!]
\includegraphics[width=1\columnwidth]{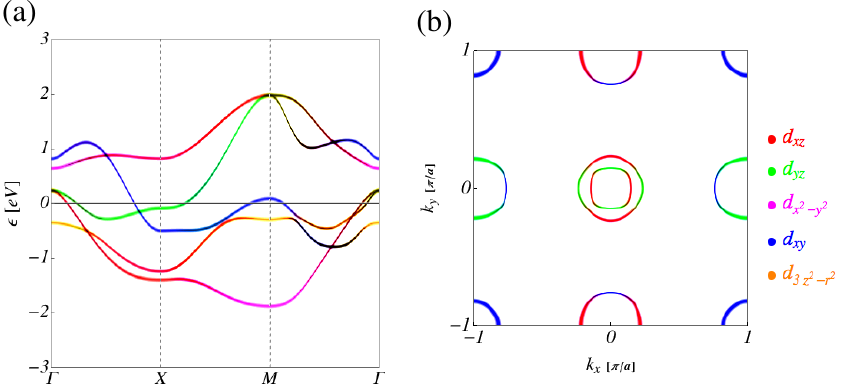}
\caption{(a) Electronic bands from Ref. \onlinecite{Ikeda2010} along the high-symmetry path $\Gamma-X-M-\Gamma$ in the 1-Fe Brillouin zone as extracted from the orbitally resolved spectral density. The zero of energy corresponds to a chemical potential realizing a filling of $n=6.0$. (b) Orbitally resolved spectral weight corresponding to the Fermi surface at band filling $ n = 6.0$.
}
\label{fig:fs}
\end{figure}

We specify the hopping parameters $t_{ij}^{\mu\nu}$ according to the bandstructure obtained by Ikeda \textit{et al.}\cite{Ikeda2010} for a five orbital model. The resulting bandstructure and orbitally resolved Fermi surface in the 1-Fe Brillouin zone is shown in Fig.~\ref{fig:fs}. The approximate nesting between hole pockets around $\Gamma$ and $M$ and electron pockets around $X$ and $Y$ promotes strong fluctuations in the particle-hole channel at wavevectors ${\bf Q}_{1} = (\pi,0)$ and ${\bf Q}_{2} = (0,\pi)$.
The leading instability in the particle-hole channel is generically found to be a spin-density wave (SDW) instability with wavevector ${\bf Q}_{1}$ or ${\bf Q}_{2}$~\cite{Chubukov2008}. The two different, orbitally resolved SDW order parameters read as ${\bf M}_{1}^{\mu\nu} = \frac{1}{\mathcal{N}} \sum_{{\bf k},\sigma,\sigma^{\prime}} \langle c_{{\bf k} + {\bf Q}_{1}\mu \sigma}^{\dagger} {\boldsymbol \sigma}_{\sigma\sigma^{\prime}} c_{{\bf k} \nu \sigma^{\prime}} \rangle$, and 
$ {\bf M}_{2}^{\mu\nu} = \frac{1}{\mathcal{N}} \sum_{{\bf k},\sigma,\sigma^{\prime}} 
\langle c_{{\bf k} + {\bf Q}_{2}\mu \sigma}^{\dagger} {\boldsymbol \sigma}_{\sigma\sigma^{\prime}} c_{{\bf k} \nu \sigma^{\prime}} \rangle, $ with $\mathcal{N}$ the number of unit cells and ${\boldsymbol \sigma}$ the vector of Pauli matrices. Taking the orbital trace yields the magnetic moments ${\bf M}_{1}$, ${\bf M}_{2}$ of the two SDW configurations.
\begin{figure}[t!]
\includegraphics[width=1\columnwidth]{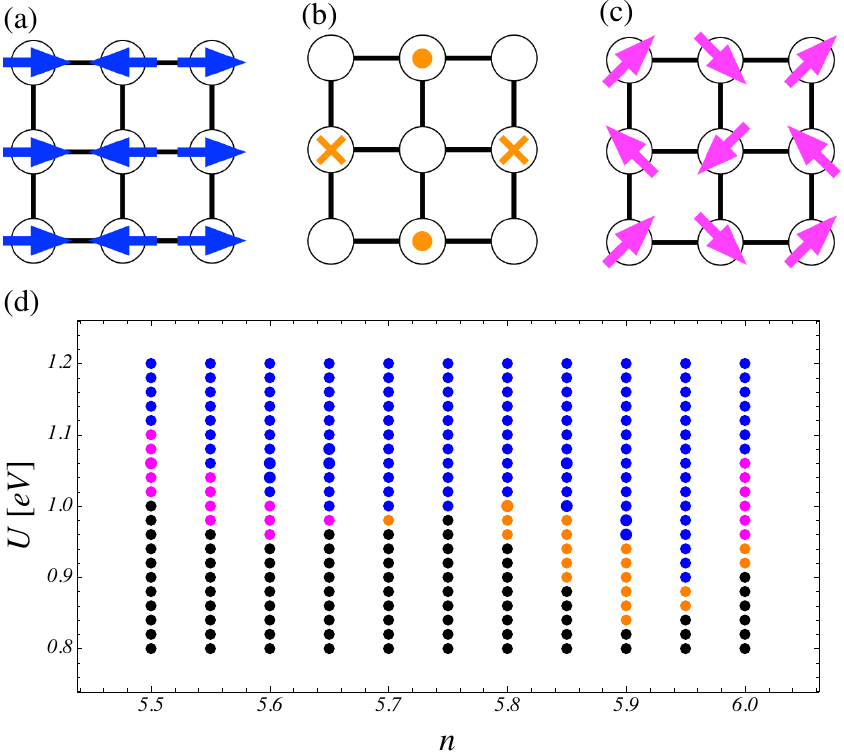}
\caption{(a) - (c) Illustration of the real space magnetization pattern of the different SDW orders, (a) magnetic stripe (MS) with ${\bf Q}_{1}$ order vector, (b) spin- and charge-ordered (SCO) state with out-of-plane moments and (c) the orthomagnetic (OM) state with non-collinear magnetic order. (d) Magnetic Hartree-Fock phase diagram obtained at $T= 0.01 \mathrm{eV}$ and Hund's coupling $J=U/4$ for the Ikeda band~\cite{Ikeda2010}. The colored symbols correspond to the paramagnetic and different magnetic states contained in our Hartree-Fock approach: ($\color{black} \bullet$) paramagnetic (PM), ($\color{blue} \bullet$)  MS, ($ \color{Orange} \bullet $) SCO and ($ \color{magenta} \bullet $) OM. In the low-$U$ region, a paramagnetic, $C_{4}$-symmetric phase is realized. As the Hubbard-$U$ is increased, a magnetic solution develops in a continuous fashion. The critical value $U_{\mathrm{c}}$ beyond which a magnetic solution is stabilized depends on the filling $n$. The magnetic order at low $U > U_{\mathrm{c}}$ is typically found to be one of the two $C_{4}$ symmetric magnetic states. As the Hubbard-$U$ is further increased, the $C_{4}$-symmetric states give way to $C_{2}$-symmetric magnetic stripes.}
\label{fig:pd}
\end{figure}
We treat interaction effects from the local Hubbard-Hund term \Eqref{eq:interaction} within a self-consistent Hartree-Fock theory adapted to the possibility of forming double-Q states~\cite{Gastiasoro2015} and describe collective fluctuations in the random-phase approximation (RPA)~\cite{Brydon2009b,Kaneshita2010,Knolle2011,Kovacic2015}, see SM~\ref{app:hartreefock},~\ref{app:hfpd} and ~\ref{app:rpa}. Within this approach the interaction both generates the commensurate SDW orders and describes the scattering of electrons with charge and magnetic fluctuations. The RPA propagators then allow us to extract the spectrum of collective excitations. An illustration of the various real-space magnetization patterns of the different SDW orders can be found in Fig.~\ref{fig:pd}(a) - (c). In Fig.~\ref{fig:pd}(d) we show a typical phase diagram in the $(n,U)$ parameter space. We find the double-Q phases to exist only at moderate values of $U$. This highlights the importance of applying an itinerant approach in the study of these phases and their collective excitations. Further computational details for the obtained phase diagram can be found in SM~\ref{app:hartreefock} and \ref{app:hfpd}. 

We further note, that the SDW ordering tendencies are strongly suppressed on the electron doped side $ n > 6 $ compared to the hole-doped side, both due to a lack of (i) nesting of electron pockets with the hole-pocket at the M-point (ii) increased distance of the Fermi level to large density-of-states contributions, see SM~\ref{app:hfpd}. Since the $C_{4}$ magnetic states exist only on the fringes of the Hartree-Fock SDW phase diagram, we expect that including the backaction of collective fluctuations on the electronic system to reduce the extent of or possibly even wipe out the $C_{4}$ phases on the electron-doped side, consistent with experimental observations.

\begin{figure*}[ht!]
\includegraphics[width=1\textwidth]{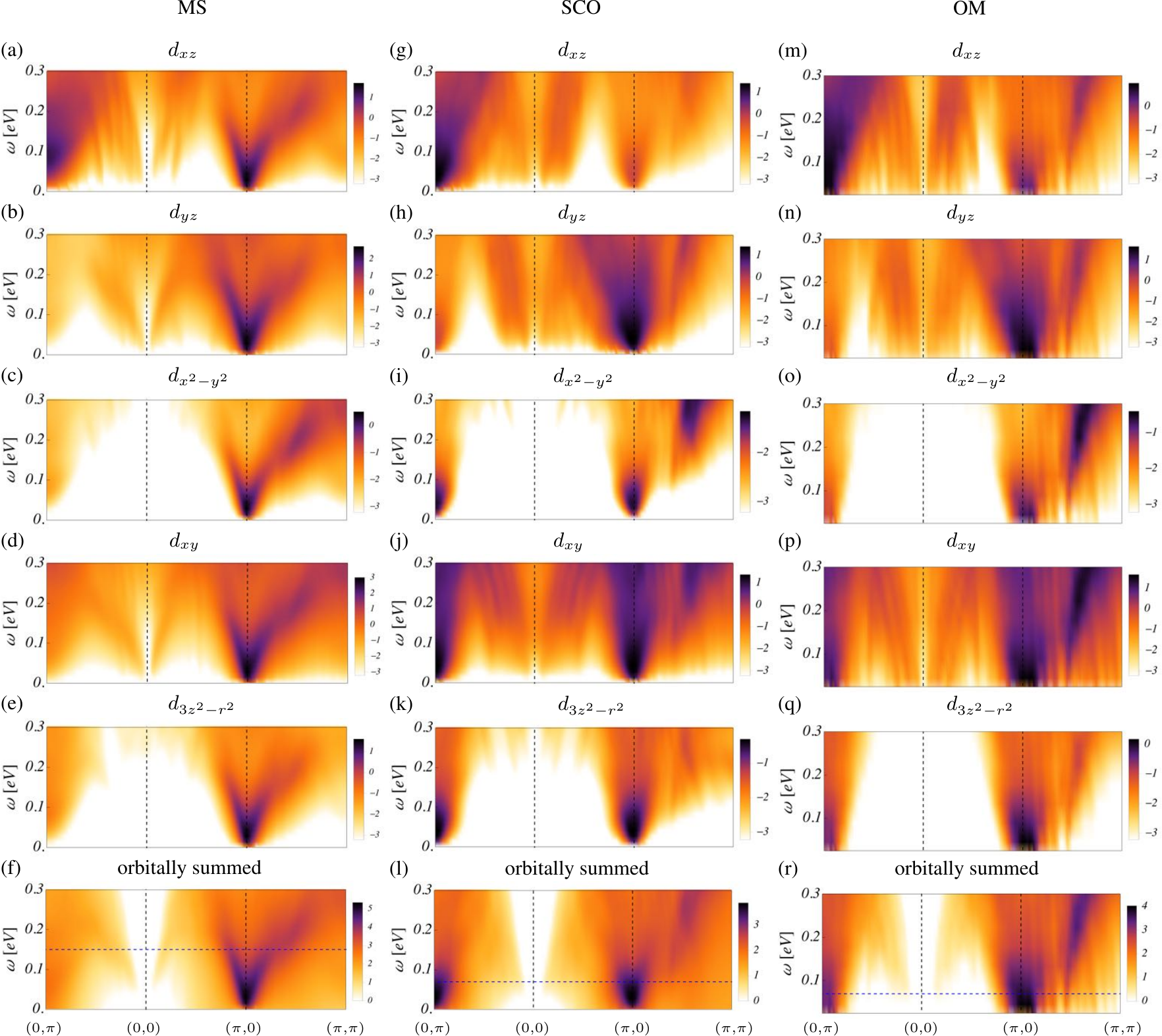}
\caption{Cut through the spectral weight distribution of magnetic excitations obtained from $\mathcal{A}^{\mu\nu}({\bf q},\omega)$ in units of $1/\mathrm{eV}$ along the high symmetry momentum-space path $(\pi,0)-(0,0)-(0,\pi)-(\pi,\pi)$. The colorscale corresponds to the log of $\mathcal{A}^{\mu\nu}({\bf q},\omega)$. The left column (a) - (f) shows the spectral weight for a (single domain) MS state stabilized for $ U = 1.05\, \mathrm{eV}$, $n = 5.87$. In (a) - (e) we display the orbitally resolved spectral weight in the intraorbital channels $\mu = \nu$. In (f) the full, orbitally summed spectral weight $\mathcal{A}({\bf q},\omega)=\sum_{\mu,\nu}\mathcal{A}^{\mu\nu}({\bf q},\omega)$ is shown. The middle column (g) - (l) shows the corresponding spectral weight of an SCO state stabilized at $ U = 0.95\, \mathrm{eV}$, $n = 5.87$, while the right column (m) - (r) shows the spectral weight for an OM state for parameters $ U = 1.02\, \mathrm{eV}$, $n = 6.00$. The blue dashed vertical lines in (f), (l), (r) mark the crossover of the gapless modes between the orbitally uniform and the orbitally polarized regime. We find the crossover scale to be at (f) $150 \, \mathrm{meV}$ (l) $ 70 \, \mathrm{meV} $, (r) $ 70 \, \mathrm{meV} $ (note the offset from $\omega = 0 $ in (r) on the vertical axis).}
\label{fig:RPA_panel}
\end{figure*}
%

\textit{Spin-excitation spectra.} 
Having stabilized the single-Q $C_{2}$ symmetric stripe SDW and the double-Q $C_{4}$ symmetric SDWs in a self-consistent manner, we are now in a position to compute the Gaussian fluctuations around these magnetic saddle-point solutions. We therefore include the Hartree-Fock self-energy $\Sigma_{\mathrm{HF}}$ in the electronic Greens function, encoding the information about the SDW states. Defining the total electron spin operator as
$
S^{i}_{{\bf q}}(\tau) = \frac{1}{\sqrt{\mathcal{N}}}\sum_{{\bf k},\mu,\sigma,\sigma^{\prime}} c_{{\bf k} + {\bf q}\mu\sigma}^{\dagger}(\tau) \frac{\sigma_{\sigma\sigma^{\prime}}^{i}}{2}c_{{\bf k}\mu\sigma^{\prime}}(\tau),
$
we compute the spin-spin correlation function for imaginary frequencies
%
%
%
\be
\label{eq:spintensor}
\chi^{ij}({\bf q},\mathrm{i}\omega_n) 
= 
\frac{1}{2\beta}\int_{0}^{\beta} \! d\tau \,
\mathrm{e}^{\mathrm{i}\omega_n \tau}
\langle \mathcal{T}_{\tau} S^{i}_{{\bf q}}(\tau) S^{j}_{-{\bf q}}(0)\rangle.
\ee 
We now express the susceptibility tensor as 
$
\chi^{ij}({\bf q},\mathrm{i}\omega_n)  = \frac{1}{4}\sum_{\mu,\nu,\{ \sigma \}}
\sigma_{\sigma_{1}\sigma_{2}}^{i}\sigma_{\sigma_{3}\sigma_{4}}^{j}
[\chi]^{\mu\sigma_1; \mu\sigma_2}_{\nu\sigma_3; \nu\sigma_4}
({\bf q},\mathrm{i}\omega_n),
$
where we define the generalized, orbitally resolved correlation function
\be 
\label{eq:orbchi}
& & \hspace{-2.25em} [\chi]^{\mu\sigma_1; \mu\sigma_2}_{\nu\sigma_3; \nu\sigma_4}({\bf q},\mathrm{i}\omega_n) = \frac{1}{2\beta\mathcal{N}}\int_{0}^{\beta} \! d\tau \,
\mathrm{e}^{\mathrm{i}\omega_n \tau} \times \\
& & \sum_{{\bf k},{\bf k}^{\prime}} \langle \mathcal{T}_{\tau} 
c_{{\bf k} + {\bf q}\mu\sigma_{1}}^{\dagger}(\tau) c_{{\bf k}\mu\sigma_{2}}(\tau)
c_{{\bf k}^{\prime} - {\bf q}\nu\sigma_{3}}^{\dagger}(0) c_{{\bf k}^{\prime}\nu\sigma_{4}}(0)
\rangle. \nn
\ee
We employ the RPA approximation to compute the generalized susceptiblity, see SM~\ref{app:rpa}. Performing the analytic continuation $\mathrm{i}\omega_{n} \to \omega + \mathrm{i}\eta$ we can extract the spectral density from the imaginary part of the retarded correlation function. We note that for the collinear MS and SCO orders, the direction of the magnetic moment, as depicted in Fig.~\ref{fig:pd}(a) - (c), defines the $\hat{\mathbf{z}}$-axis of our coordinate system in spin space. In this case, a separation of magnetic excitations into transverse and longitudinal with respect to the orientation of the ordered moment is possible. This is, however, not the case for the non-collinear OM state.

We note that the advantage of using the RPA type approach in application to the itinerant magnets is the immediate information on the collective excitations, associated with the magnetic ordering, as well as particle-hole continuum of the gapless fermionic modes and their interaction with the former. Furthermore, while the full susceptibility tensor is orbitally summed, we also naturally have access to the orbitally resolved contributions to the total susceptibility \Eqref{eq:spintensor} through the generalized correlation function \Eqref{eq:orbchi}. In particular, we can analyze the orbital structure of magnetic excitations. This provides valuable insight both for experiment and for the modelling of low-energy Hamiltonians that aim at describing the (undamped) magnetic excitations of the FeSC parent materials.

We find that the most prominent differences between the spin excitations of the $C_{2}$ and $C_{4}$ phases are their respective orbital polarizations outside of the immediate vicinity of the ordering wave vector. To compare between excitation spectra of collinear and non-collinear magnetic orders, we study the spectral function
\be 
\label{eq:spectral}
\mathcal{A}^{\mu\nu}({\bf q},\omega) = \mathrm{Im}\sum_{i,j,\{ \sigma \}}\sigma_{\sigma_{1}\sigma_{2}}^{i}\sigma_{\sigma_{3}\sigma_{4}}^{j} 
[\chi]^{\mu\sigma_1; \mu\sigma_2}_{\nu\sigma_3; \nu\sigma_4}({\bf q},\omega), 
\ee
that contains contributions from all spin-channels. In the case of collinear SDW order (MS, SCO) only the $i = j$ terms corresponding to transverse and longitudinal spin excitations are finite, while $i \neq j$ vanishes by symmetry. In the non-collinear case (OM) also off-diagonal terms contribute. The qualitative properties of the spin-susceptibility tensor reported for the local-moment approach~\cite{Wang2015} are reproduced within our itinerant scenario, as is also confirmed by the number of soft modes. For MS and SCO, there exist two Goldstone modes due to spin-rotational symmetry breaking, while the OM phase is characterized by three gapless collective excitations. We note that the number of gapless modes is not affected by the additional orbital structure, as can be verified by a full diagonalization of~\Eqref{eq:spintensor} in spin \textit{and} orbital space.   

In Fig.~\ref{fig:RPA_panel} we show our main results for the orbitally resolved spin-excitation spectra extracted from~\Eqref{eq:spectral}. We focus on the intraorbital contributions and first analyze the properties of the magnetic excitations in the stripe state as shown in Fig.~\ref{fig:RPA_panel}(a)-(f). We can clearly identify the (twofold degenerate) soft mode in the excitation spectra emerging around the ordering vector ${\bf Q}_{1}$ of the stripe state. The spin-excitation spectrum is not $C_{4}$ symmetric due to the $C_{4}$ to $C_{2}$ symmetry breaking of the single-Q state. The excitation spectrum features rather sharp and well-defined spinwave branches up to $\sim 150\,\mathrm{meV}$. A detailed study of the spinwave anisotropy of the MS state and the evolution of the itinerant multiorbital scenario to the Heisenberg scenario of local moments was performed in Ref.~\onlinecite{Kovacic2015}. While the spectral weight at zero transfer momentum is completely suppressed, gapped features are visible at the ordering vector ${\bf Q}_{2} $ of the degenerate stripe-ordered state. The longitudinal excitations in the $\mu = \nu = d_{xz}$ channel contribute most weight to the gapped spectral feature at ${\bf Q}_{2}$. The dominating, coherent spinwave branches on the other hand show a robust orbital uniformity in the sense that the shape of the spinwave dispersion is the same for all intraorbital channels for excitation energies up to roughly $\sim 150\,\mathrm{meV}$. Above this energy scale, the excitations become increasingly incoherent and show an enhanced orbital polarization SM~\ref{app:orbpol}.

In contrast to this, the magnetic excitations of the SCO state, as displayed in Fig.~\ref{fig:RPA_panel}(g)-(l), feature a strong orbital dependence in the $d_{xz}$, $d_{yz}$ manifold already at low energies. In fact, the intensity in the intraorbital $d_{yz}$ / $d_{xz}$ component is dominated by fluctuations at ${\bf Q}_{1}$ / ${\bf Q}_{2}$. The degree of orbital polarization actually decreases with increasing energy above $\sim 70 \, \mathrm{meV}$ and momenta deviating from the ordering vectors, see SM~\ref{app:orbpol}. Instead of dispersive spinwave branches, the spectral distribution of magnetic excitations features broad, pillar-like structures. The absence of well-defined spinwave branches can be understood from a small SDW gap and a concurrent damping of the magnetic excitations due to the presence of the particle-hole continuum. The orbitally summed spectral weight has soft modes around both ordering vectors ${\bf Q}_{1}$ and ${\bf Q}_{2}$. Naturally, the excitation spectrum around the $C_{4}$ symmetric SCO state is fully $C_{4}$ symmetric. A $C_{4}$ rotation, maps the intraorbital $d_{xz}$ and $d_{yz}$ spectra onto each other, as the orbital degrees of freedom participate in spatial rotations. Although the SCO state in principle allows for charge order, we found the ${\bf Q}_{3}$ charge modulation to be almost negligible featuring a gapped spectrum for collective excitations. 

Our results for the OM state are collected in Fig.~\ref{fig:RPA_panel}(m)-(r). While certain features distinguish the coplanar OM state from the collinear SCO state, the orbitally resolved and orbitally summed magnetic excitation spectra are very similar between the two cases and feature the same type of orbital polarization of the excitations in the $d_{xz}$, $d_{yz}$ manifold.
We note that, as the computation of the spin-spin correlation tensor is numerically more demanding than the separate computation of transverse and longitudinal fluctuations for the MS and SCO cases, we employed a coarser momentum-frequency grid to produce the results in Fig.~\ref{fig:RPA_panel}(m)-(r). 

Another important observation for the collective excitations of the C$_4$ symmetric phases is that their collective spin excitations cannot be mapped on the effective low-energy spin-only models at any energy ranges. In particular, while for the $C_2$ symmetric striped antiferromagnetic state one may in principle fit the low-energy part of the spin excitation spectrum, obtained by us, using the effective $J_1-J_2$ model with biquadratic exchange, this approach would completely fail in the C$_4$ symmetric OM or SCO phases. This points towards purely itinerant nature of these phases which should be clearly seen in the experimental data. In particular, our spin excitations for the OM phase strongly differ from the spin waves obtained within the localized description used recently\cite{Wang2015}.

In conclusion, our results for the magnetic excitations of the single-Q $C_{2}$ symmetric and double-Q $C_{4}$ symmetric magnetic states for the FeSC suggest a strong orbital polarization for the latter, that can in principle distinguish double-Q from twinned single-Q states in orbitally resolved measurements of the excitation spectra. One of the potential ways to do that is to measure magnetic excitations using RIXS technique.  In the case of twinned MS states, exciting in the channel with $d_{yz}$ orbital symmetry will create a gapless $d_{yz}$ signal around ${\bf Q}_{1}$ and its twin around ${\bf Q}_{2}$. In a $C_{4}$ symmetric state, however, the same set up will create only one gapless feature around ${\bf Q}_{1}$. We further conclude from the observed orbital uniformity of excitations of the MS state, that its qualitative features can indeed be modelled by low-energy Heisenberg-type Hamiltonians. We propose, however, that such a description will fail to capture the properties of the $C_{4}$ magnetic states, and the Heisenberg-Hamiltonian for spin degrees of freedom has to be supplemented by operators acting in the orbital $d_{xz}$, $d_{yz}$ subspace. Finally, the spin excitations of the $C_4$ symmetric phases have purely itinerant character, which cannot be described within the localized type of model. Thus an experimental measurement of the spin excitation spectra in the SCO phase observed recently in the hole-doped iron pnictides should provide a clear hallmark of itineracy of the collective excitations in the FeSC.

We thank M. H. Christensen, R. M. Fernandes, M. N. Gastiasoro, and A. Kreisel for useful discussions. D.D.S. acknowledges support from the Villum foundation. B.M.A, acknowledges support from Lundbeckfond fellowship (grant A9318). 



\newpage


\begin{widetext}

\setcounter{equation}{0}
\setcounter{figure}{0}
\setcounter{table}{0}
\setcounter{page}{1}
\makeatletter
\renewcommand{\theequation}{S\arabic{equation}}
\renewcommand{\thefigure}{S\arabic{figure}}
\renewcommand{\bibnumfmt}[1]{[S#1]}
\renewcommand{\citenumfont}[1]{S#1}

\begin{center}
\textbf{\large Supplementary Material: ``Collective magnetic excitations of $C_{4}$ symmetric magnetic states in iron-based superconductors''}
\end{center}

\section{Hartree-Fock Hamiltonian}
\label{app:hartreefock}

To study commensurate magnetic order and magnetic excitations in magnetically ordered states of parent materials of FeSC, we perform a Hartree-Fock decoupling of the Hubbard-Hund interaction \Eqref{eq:interaction} as was done in Ref. \onlinecite{Gastiasoro2015}. We include the ordering vectors ${\bf Q}_{1} = (\pi, 0)$ and 
${\bf Q}_{2} = (0,\pi)$ corresponding to the two stripy SDW components ${\bf M}_{1}$ and ${\bf M}_{2}$. To allow for a description of double-Q phases, we also include the ordering vector ${\bf Q}_{3} = (\pi,\pi)$.

For each Bloch vector ${\bf k}$ the Hartree-Fock Hamiltonian is then specified by a $40 \times 40$ matrix in spin $\otimes$ orbital $\otimes$ Umklapp space. The self-consistency
equations can be concisely stated as
\be
\label{eq:mf1}
n_{0}^{\mu\nu} & =  & \frac{1}{\mathcal{N}} \sum_{{\bf k}, \sigma} 
\langle c_{{\bf k} \mu \sigma}^{\dagger} c_{{\bf k} \nu \sigma} \rangle, 
\quad
n_{3}^{\mu\nu}  = \frac{1}{\mathcal{N}} \sum_{{\bf k}, \sigma} 
\langle c_{{\bf k} + {\bf Q}_{3} \mu \sigma}^{\dagger} c_{{\bf k} \nu \sigma} \rangle, \\
M_{1}^{\mu\nu}  & = & \frac{1}{\mathcal{N}} \sum_{{\bf k}, \sigma}\sigma
\langle c_{{\bf k} + {\bf Q}_{1} \mu \sigma}^{\dagger} c_{{\bf k} \nu \sigma} \rangle, 
\quad
M_{2}^{\mu\nu}  =  \frac{1}{\mathcal{N}} \sum_{{\bf k}, \sigma}\sigma
\langle c_{{\bf k} + {\bf Q}_{2} \mu \sigma}^{\dagger} c_{{\bf k} \nu \sigma} \rangle, 
\ee
and for the spin off-diagonal fields
\be
\label{eq:mf3}
\tilde{n}_{0}^{\mu\nu} & = & \frac{1}{\mathcal{N}} \sum_{{\bf k}, \sigma}\sigma
\langle c_{{\bf k} \mu \bar{\sigma}}^{\dagger} c_{{\bf k} \nu \sigma} \rangle, 
\quad
\tilde{n}_{3}^{\mu\nu} = \frac{1}{\mathcal{N}} \sum_{{\bf k}, \sigma}\sigma 
\langle c_{{\bf k} + {\bf Q}_{3} \mu \bar{\sigma}}^{\dagger} c_{{\bf k} \nu \sigma} \rangle, 
\\
\label{eq:mf4}
\tilde{M}_{1}^{\mu\nu} & = & \frac{1}{\mathcal{N}} \sum_{{\bf k}, \sigma}
\langle c_{{\bf k} + {\bf Q}_{1} \mu \bar{\sigma}}^{\dagger} c_{{\bf k} \nu \sigma} \rangle, 
\quad
\tilde{M}_{2}^{\mu\nu} = \frac{1}{\mathcal{N}} \sum_{{\bf k}, \sigma}
\langle c_{{\bf k} + {\bf Q}_{2} \mu \bar{\sigma}}^{\dagger} c_{{\bf k} \nu \sigma} \rangle,
\ee
with $\mathcal{N}$ the number of unit cells and the ${\bf k}$-sum extends over the full Brillouin zone corresponding to the 1-Fe unit cell. The matrix-valued mean-field $n_{0}^{\mu\nu} $ renormalizes the chemical potential, but also allows for 
orbital-dependent shifts. The mean-field $n_{3}^{\mu\nu}$ describes a charge-density wave state with a $(\pi,\pi)$ modulation of the charge distribution. The mean-fields
$M_{1}^{\mu\nu}$, $\tilde{M}_{1}^{\mu\nu} $ and $M_{2}^{\mu\nu}$, 
$\tilde{M}_{2}^{\mu\nu} $ completely specify the $x$ and $z$ components of the SDW order-parameters ${\bf M}_{1}$ and ${\bf M}_{2}$. Due to the specific choice of mean-field assumptions that correspond to fixing a spin-quantization axis, the $y$ components of the stripy SDW order-parameters vanishes identically. Finally,  $\tilde{n}_{0}^{\mu\nu}$ and $\tilde{n}_{3}^{\mu\nu}$ describe ferromagnetic and antiferromagnetic N$\acute{\textrm{e}}$el order, respectively. By virtue of our mean-field assumptions, they can only acquire finite values in the $y$ direction. 

The average $\langle \cdots \rangle$ on the right hand side is computed with respect to a thermal state of the Hartree-Fock Hamiltonian $H_{\mathrm{HF}} = \sum_{{\bf k},\mu,\nu\sigma}^{\prime}\Psi_{{\bf k}\mu}^{\dagger}h^{\mu\nu}({\bf k})\Psi_ {{\bf k}\nu}$. The Bloch-Hamiltonian $h^{\mu\nu}({\bf k})$ containing the mean-fields Eqs. (\ref{eq:mf1})-(\ref{eq:mf4}) is defined with respect to the reduced Brillouin zone $ [-\pi/2,\pi/2) \times [-\pi/2,\pi/2)$. It can be written as
\begin{eqnarray}
h^{\mu\nu}({\bf k}) = 
\begin{pmatrix}
 \xi^{\mu\nu}({\bf k})& W^{\mu\nu}_1 & W^{\mu\nu}_2 &N^{\mu\nu}_3 &\tilde N^{\mu\nu}_0 &\tilde W^{\mu\nu}_1 &\tilde W^{\mu\nu}_2 &\tilde N^{\mu\nu}_3 \\
 +N^{\mu\nu}_0& && & & & & \\
 &\xi^{\mu\nu}({\bf k}+{\bf Q}_1) &N^{\mu\nu}_3 &W^{\mu\nu}_{2}&\tilde W^{\mu\nu}_{1} & \tilde N^{\mu\nu}_0&\tilde N^{\mu\nu}_3 &\tilde W^{\mu\nu}_2 \\
 &+N^{\mu\nu}_0 & & & & & & \\
 & & \xi^{\mu\nu}({\bf k}+{\bf Q}_2)&W^{\mu\nu}_{1}  & \tilde W^{\mu\nu}_{2}&\tilde N^{\mu\nu}_3 &\tilde N^{\mu\nu}_0 &\tilde W^{\mu\nu}_1 \\
 & &+N^{\mu\nu}_0 & & & & & \\
 & & &\xi^{\mu\nu}({\bf k}+{\bf Q}_3)&\tilde N^{\mu\nu}_3 & \tilde W^{\mu\nu}_{2}& \tilde W^{\mu\nu}_{1}&\tilde N^{\mu\nu}_0 \\
 & & &+N^{\mu\nu}_0 & & & & \\
 & & & & \xi^{\mu\nu}({\bf k})& -W^{\mu\nu}_1&-W^{\mu\nu}_2 &N^{\mu\nu}_3 \\
 & & & &+N^{\mu\nu}_0 & & & \\
 & & & & & \xi^{\mu\nu}({\bf k}+{\bf Q}_1)& N^{\mu\nu}_3&-W^{\mu\nu}_{2} \\
 & & & & &+N^{\mu\nu}_0 & & \\
 & & \mathrm{h.c.} & & & &\xi^{\mu\nu}({\bf k}+{\bf Q}_2)&-W^{\mu\nu}_{1} \\
 & & & & & &+N^{\mu\nu}_0 & \\
 & & & & & & &\xi^{\mu\nu}({\bf k}+{\bf Q}_3)\\ 
 & & & & & & &+N^{\mu\nu}_0 \\
\end{pmatrix}, \nn
\end{eqnarray}
where the basis is defined by the spinor
\begin{eqnarray}
\Psi_{{\bf k}\nu}^{\dagger}=
\begin{pmatrix}
c_{{\bf k}\mu\uparrow}^{\dagger} &
c_{{\bf k} + {\bf Q}_{1}\mu\uparrow}^{\dagger} &
c_{{\bf k} + {\bf Q}_{2}\mu\uparrow}^{\dagger} &
c_{{\bf k} + {\bf Q}_{3}\mu\uparrow}^{\dagger} &
c_{{\bf k}\mu\downarrow}^{\dagger} &
c_{{\bf k} + {\bf Q}_{1}\mu\downarrow}^{\dagger} &
c_{{\bf k} + {\bf Q}_{2}\mu\downarrow}^{\dagger} &
c_{{\bf k} + {\bf Q}_{3}\mu\downarrow}^{\dagger}
 \end{pmatrix}.
\end{eqnarray}
The orbital matrices $N_{0}^{\mu\nu}$, $N_{3}^{\mu\nu}$, $\tilde{N}_{0}^{\mu\nu}$, $\tilde{N}_{3}^{\mu\nu}$ and $W_{1}^{\mu\nu}$, $W_{2}^{\mu\nu}$, $\tilde{W}_{1}^{\mu\nu}$, $\tilde{W}_{2}^{\mu\nu}$ entering $h^{\mu\nu}({\bf k})$ are composed of the charge and magnetic mean-fields in Eqs. (\ref{eq:mf1})-(\ref{eq:mf4}) as
\be
\label{eq:N0}
N_{0}^{\mu\nu} & = & 
\delta^{\mu\nu}\Bigl(U n_{0}^{\mu} + (2 U^{\prime} - J) \bar{n}_{0}^{\nu}\Bigr) + 
\bar{\delta}^{\mu\nu} 
\Bigl((-U^{\prime} + 2 J) n_{0}^{\nu\mu} + J^{\prime} n_{0}^{\mu\nu} \Bigr), \\
\label{eq:N3}
N_{3}^{\mu\nu} & = & \delta^{\mu\nu}\Bigl( U n_{3}^{\mu} + (2 U^{\prime} - J)\bar{n}_{3}^{\nu} \Bigr) +
\bar{\delta}^{\mu\nu} \Bigl( (-U^{\prime} + 2 J) n_{3}^{\nu\mu}+ J^{\prime}n_{3}^{\mu\nu} \Bigr), \\
\label{eq:W1}
W_{1}^{\mu\nu} & = & \delta^{\mu\nu}\Bigl(-U M_{1}^{\mu} - J \bar{M}_{1}^{\nu} \Bigr) + \bar{\delta}^{\mu\nu}
\Bigl( U^{\prime} M_{1}^{\nu\mu} - J^{\prime} M_{1}^{\mu\nu} \Bigr), \\
\label{eq:W2}
W_{2}^{\mu\nu} & = & \delta^{\mu\nu}\Bigl(-U M_{2}^{\mu} - J \bar{M}_{2}^{\nu} \Bigr) + \bar{\delta}^{\mu\nu}
\Bigl( U^{\prime} M_{2}^{\nu\mu} - J^{\prime} M_{2}^{\mu\nu} \Bigr) 
\ee
and
\be
\label{eq:NT0}
\tilde{N}_{0}^{\mu\nu} & = & \delta^{\mu\nu} \Bigl( -U \tilde{n}_{0}^{\mu} - J \bar{\tilde{n}}_{0}^{\nu} \Bigr) + \bar{\delta}^{\mu\nu} \Bigl( -U^{\prime} \tilde{n}_{0}^{\nu\mu} - J^{\prime}\tilde{n}_{0}^{\mu\nu} \Bigr), \\
\label{eq:NT3}
\tilde{N}_{3}^{\mu\nu} & = & \delta^{\mu\nu} \Bigl(-U \tilde{n}_{3}^{\mu} - J \bar{\tilde{n}}_{3}^{\nu} \Bigr) +
\bar{\delta}^{\mu\nu}\Bigl( -U^{\prime} \tilde{n}_{3}^{\nu\mu} - J^{\prime} \tilde{n}_{3}^{\mu\nu} \Bigr), \\
\label{eq:WT1}
\tilde{W}_{1}^{\mu\nu} & = & \delta^{\mu\nu}\Bigl( -U \tilde{M}_{1}^{\mu} - J \bar{\tilde{M}}_{1}^{\nu} \Bigr) +
\bar{\delta}^{\mu\nu}\Bigl( -U^{\prime} \tilde{M}_{1}^{\nu\mu} - J^{\prime} \tilde{M}_{1}^{\mu\nu} \Bigr), \\
\label{eq:WT2}
\tilde{W}_{2}^{\mu\nu} & = & \delta^{\mu\nu}\Bigl( -U \tilde{M}_{2}^{\mu} - J \bar{\tilde{M}}_{2}^{\nu} \Bigr) +
\bar{\delta}^{\mu\nu}\Bigl( -U^{\prime} \tilde{M}_{2}^{\nu\mu} - J^{\prime} \tilde{M}_{2}^{\mu\nu} \Bigr). 
\ee
Following the notation in Ref.~\onlinecite{Gastiasoro2015} we have introduced further
auxiliary quantities to ease the notation, where $\delta^{\mu\nu}$ denotes the Kronecker symbol with respect to orbital indices and $\bar{\delta}^{\mu\nu} = 1 - \delta^{\mu\nu}$ filters out the orbital off-diagonal components. We note, that repeated indices are \emph{not} summed over. Quantities in Eqs.~(\ref{eq:N0})-(\ref{eq:WT2}) with a single orbital index refer to the diagonal element of the corresponding matrix, e.g. $ n_{0}^{\mu} = n_{0}^{\mu\mu} $. Objects with a bar, such as $\bar{n}_{0}^{\nu}$, are defined as, e.g., $\bar{n}_{0}^{\nu} = \sum_{\mu \neq \nu} n_{0}^{\mu\mu}$. The bare dispersion enters through 
$\xi^{\mu\nu}({\bf k}) = \varepsilon^{\mu\nu}({\bf k}) -   \delta^{\mu\nu} \mu_{0}$, where $\epsilon^{\mu\nu}({\bf k}) $ is obtained from the Bloch representation of the hopping Hamiltonian \Eqref{eq:hopping} and $\mu_{0} $ is the chemical potential controlling the filling of the electronic bands.

The self-consistent set of equations Eqs.~(\ref{eq:mf1})-(\ref{eq:mf4}) for the charge and magnetic order-parameters, supplemented by fixing the filling of the electronic system through the self-consistent determination of the chemical potential, is then solved iteratively starting with a random seed. While in this momentum space formulation, rather large system sizes are in principle easily accessible, we stabilize magnetic states for moderate system sizes for practical reasons. The computation of the susceptibility in the double-Q phases increases the dimensionality of the susceptibility matrices (see App.~\ref{app:rpa}) and correspondingly increases the numerical effort. The results presented in this work have been obtained for a 10 $\times$ 10 mesh in the reduced Brillouin zone, which corresponds to a 20 $ \times $ 20 mesh in the original 1-Fe Brillouin zone.

\section{Hartree-Fock phase diagram}
\label{app:hfpd}

We solved the set of self-consistent equations Eqs.~(\ref{eq:mf1})-(\ref{eq:mf4}) numerically by starting with a random initial seed and iterating to self-consistency. In the following we restrict our attention to the temperature $k_{\mathrm{B}}T = 0.01 \, \mathrm{eV}$ and sweep a rather wide filling range of $ n = 5.5, \dots, 6.0 $ and interaction parameters $ U = 0.8, \dots, 1.2 \, \mathrm{eV}$, $J = U/4$. 

The temperature vs. filling phase diagram was studied in detail for a certain set of interaction parameters previously ~\cite{Gastiasoro2015} and we find agreement with these results where a direct comparison is available. Our phase diagram in the filling vs. Hubbard-$U$ parameter space is summarized in Fig.~\ref{fig:pd}. A decrease of the Hund's coupling below the value of $U/4$ will eventually shift the phase diagram upward on the $U$-axis and the phase boundaries steepen up. The global phase structure, however, remains largely the same. The onset of magnetic order is typically seen as a continuous growth in the magnetic moment as the Hubbard-$U$ is increased. For $J = U/4$, the phase boundary separating the paramagnetic from the magnetic phase occurs at finite value of the $U$. The phase-boundary itself is somewhat sensitive to the changes in the doping level within certain bounds. Within the doping range $  n = 5.5, \dots, 6.0 $ we find the lowest critical value $U_{\mathrm{c}}(n)$ at $ n = 5.9 $. This appears consistent with previous results~\cite{Gastiasoro2015}, where the largest SDW transition temperature for the employed hopping parameters~\cite{Ikeda2010} was obtained for a filling $ n = 5.91 $. 
\begin{figure*}[ht!]
\includegraphics[width=1\columnwidth]{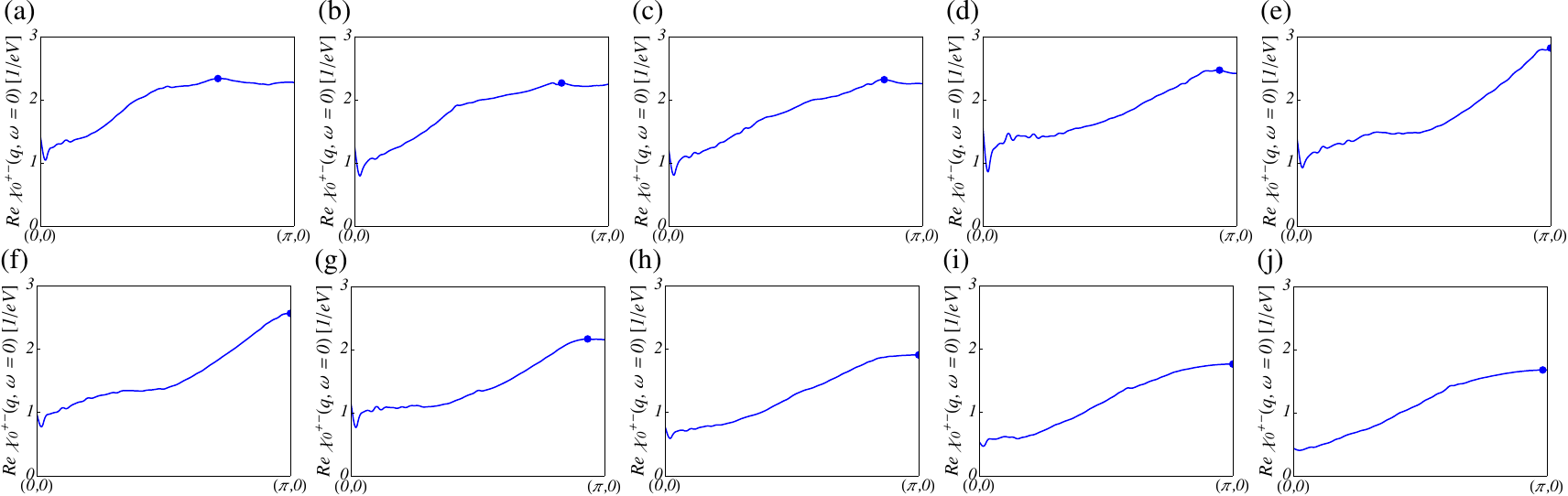}
\caption{Real part of $\chi_{0}^{+-}({\bf q}, \omega = 0)$ along the cut $(0,0) - (\pi,0)$ in momentum space. (a) - (e) Evolution of the bare susceptibility (without including interaction effects) for fillings (a) $ n = 5.5 $, (b) $ n = 5.6 $, (c) $ n = 5.7$, (d) $ n = 5.8 $, (e) $n = 5.9$ on the hole-doped side. (f) - (j) The same for fillings (f) $ n = 6.0 $, (g) $ n = 6.1 $, (h) $ n = 6.2 $, (i) $ n = 6.3 $, (j) $n = 6.4 $ including the electron-doped side. The blue dot marks the respective maximum in the real part of the susceptibility. While the degree of incommensurability is stronger on the hole-doped side, in comparison the overall scale of the susceptibility is decreased on the electron doped side.}
\label{fig:suppsusc}
\end{figure*}
For $ n > 6 $, the critical value of the onsite interaction required to push the system into the SDW state increases strongly with increased electron doping. We also observe that the region where one can find a dome-shaped SDW phase in the temperature vs. doping parameter space in a reasonable doping range is restricted to moderate values of the Hubbard interaction. The different behavior of ordering tendencies on hole- and electron-doped sides can be understood from the bare susceptibility $\chi_{0}^{+-}({\bf q}, \omega = 0)$ in the paramagnetic phase. In Fig.~\ref{fig:suppsusc}, we display the evolution of the bare susceptibility along the $(0,0) - (\pi,0)$ cut in momentum space with changing the filling from hole doped to electron doped. In both cases, incommensurate tendencies appear, albeit in a non-monotonuous way on the electron-doped side. Furthermore observe the overall factor of 2 reduction of the peak height at ${\bf q} = (\pi,0)$ in $\mathrm{Re} \chi_{0}^{+-}({\bf q}, \omega = 0)$ for the electron doping as compared to the hole-doped counterpart.

We find that most part of the parameter space where the free-energy displays a magnetic solution is taken up by the single-Q stripe phase with ordering vector either ${\bf Q}_{1}$ or ${\bf Q}_{2}$. The magnetic mean-fields are dominated by intraorbital contributions, and off-diagonal contributions emerge only for the $d_{3z^{2}-r^{2}}$ and $d_{x^2-y^{2}}$ orbitals. The magnetic moment increases monotonously as a function of $U$. The $3d$-Fe orbitals contribute differently to the magnetic order parameter. In the stripe phase, either the intraorbital $d_{yz}$ or the $d_{xz}$ contribution to ${\bf M}_{1}^{\mu\nu}$ or ${\bf M}_{2}^{\mu\nu}$ dominates, depending on whether the ${\bf Q}_{1}$ or the ${\bf Q}_{2}$ stripe SDW is realized. The $d_{xy}$ orbital can provide a contribution of similar size, as does the $d_{3z^{2}-r^{2}}$. The $d_{x^{2}-y^{2}}$ contribution is typically an order of magnitude smaller. The precise orbital composition is, however, sensitive to changes in filling and interaction parameters.

For our choice of the Hund's coupling $J=U/4$, we find the double-Q phases where both SDWs with ${\bf Q}_{1}$ and ${\bf Q}_{2}$ contribute simultaneously in a coherent fashion to occur more or less exclusively in the low-$U$ region of the phase diagram. Interestingly, the transition from the $ C_{4} $ symmetric paramagnetic phase to a magnetic state occurs through the $ C_{4} $ symmetric double-Q SDWs in the entire filling range considered. The SCO state can be stabilized for $ n = 5.7 $ to $ n = 6.0 $ with a small $n_{3}^{\mu\nu}$ component describing the checkerboard charge order with ordering vector $ {\bf Q}_{3} $. The hierarchy of the orbital composition entering the magnetic order parameters is the same as for the stripy state. But in accord with the $C_{4}$ symmetry, the intraorbital components satisfy ${\bf M}_{1}^{xz,xz} = \pm {\bf M}_{2}^{yz,yz}$ and ${\bf M}_{1}^{yz,yz} = \pm {\bf M}_{2}^{xz,xz}$, where $\pm $ corresponds to parallel/antiparallel orientation of the magnetic moments. The remaining intraorbital contributions are the same for both SDW components, while the off-diagonal term mixing $d_{3z^{2}-r^{2}}$ and $d_{x^{2}-y^{2}}$ has a relative sign between the ${\bf Q}_{1}$ and the ${\bf Q}_{2}$ SDW component. The interaction controlled magnetic moment ranges from $M \sim 0.1 \, \mu_{\mathrm{B}}$ to up to $M \sim 0.5 \, \mu_{\mathrm{B}}$ within the SCO phase.

The extent of the SCO phase in the $U$-direction narrows down as the outer parts of the SCO region are approached by changes in the filling. At the boundaries of the SCO phase in doping direction, our Hartree-Fock solver converges to OM states with non-collinear magnetic order. The SCO state is thus enclosed by two OM phases. They occur for slightly larger interaction strengths and are stable up to larger interactions than the SCO state. We find the same overall hierarchy in the orbital composition of the OM state as for the SCO state. The SDW order-parameters are now related by e.g. $ M_{1}^{yz,yz} = \tilde{M}_{2}^{xz,xz}$ and $\tilde{M}_{1}^{yz,yz} = -M_{2}^{xz,xz}$ (the same relation holds for $xz \leftrightarrow yz $), $ M_{1}^{\mu\mu} = \tilde{M}_{2}^{\mu\mu}$ and $ \tilde{M}_{1}^{\mu\mu} = -M_{2}^{\mu\mu}$ for the remaining orbitals. For the only non-zero off-diagonal elements mixing the $d_{3z^{2}-r^{2}}$ and $d_{x^{2}-y^{2}}$ orbital, one has $ M_{1}^{\mu\nu} = -\tilde{M}_{2}^{\mu\nu}$ and $ \tilde{M}_{1}^{\mu\nu} = M_{2}^{\mu\nu}$. We note, that degenerate OM states with modified relations between the SDW components exist. The magnetic moment can reach values of $m \sim 1.4 \, \mu_{\mathrm{B}}$ in the OM phase. As $U$ increases beyond $ U \sim 1 \, \mathrm{eV} $, the SCO and OM states cease to be stable and give way to stripy states.

\section{RPA equations without spin-rotation symmetry}
\label{app:rpa}

Here we briefly describe the RPA formalism we employ to compute the
spin-excitations in the magnetically ordered phases. For the sake of generality,
we assume (i) the absence of spin-rotation symmetry and (ii) the presence of double-Q
magnetic order encoded in the electronic self-energy. The electronic self-energy that describes the magnetic order is obtained form the self-consistent Hartree-Fock procedure outlined in 
App.~\ref{app:hartreefock}. The matrices $N_{0}^{\mu\nu}$, $N_{3}^{\mu\nu}$, $\tilde{N}_{0}^{\mu\nu}$, $\tilde{N}_{3}^{\mu\nu}$ and $W_{1}^{\mu\nu}$, $W_{2}^{\mu\nu}$, $\tilde{W}_{1}^{\mu\nu}$, $\tilde{W}_{2}^{\mu\nu}$ define our static approximation for the electronic self-energy, i.e., $\Sigma^{\mu\nu}({\bf k},\mathrm{i}\omega_{n}) = \Sigma_{\mathrm{HF}}^{\mu\nu}$  . While in a paramagnetic states or a state with
collinear magnetic order, the conservation of the $z$-component of the electronic spin facilitates a decoupling of the RPA equations for transverse and longitudinal fluctuations, this is no longer the case in a non-collinear magnetic state. While stripe order and the spin- and charge-ordered state realize a collinear magnetic structure, the orthomagnetic state is non-collinear.

To make the connection to neutron-scattering experiments, we compute the imaginary-time spin-spin correlation function
\be
\chi^{ij}({\bf q},\mathrm{i}\omega_n) = \frac{1}{2\beta}\int_{0}^{\beta} \! d\tau \,
\mathrm{e}^{\mathrm{i}\omega_n \tau}
\langle \mathcal{T}_{\tau} S^{i}_{{\bf q}}(\tau) S^{j}_{-{\bf q}}(0)\rangle 
\ee
with the electron spin operators defined as
\be
S^{i}_{{\bf q}}(\tau) = \frac{1}{\sqrt{\mathcal{N}}}\sum_{{\bf k},\mu,\sigma,\sigma^{\prime}} c_{{\bf k} + {\bf q}\mu\sigma}^{\dagger}(\tau) \frac{\sigma_{\sigma\sigma^{\prime}}^{i}}{2}c_{{\bf k}\mu\sigma^{\prime}}(\tau).
\ee
Here $\mathcal{T}_{\tau}$ denotes the time-ordering operator with respect to the imaginary-time variable $\tau \in [0,\beta)$, with $\beta$ the inverse temperature, and $i,j = x,y,z$ label the spatial components of the spin operator with respect to a given coordinate system, with $\sigma_{\sigma\sigma^{\prime}}^{i}$ the $i$-th Pauli matrix.

From the imaginary part of $\chi^{ij}({\bf q},\mathrm{i}\omega_n)$, we can extract
the spectrum of spin-excitations that are probed by neutron scattering. In our approach that we
will outline below, also density-density correlations are easily accessible. The density susceptibility is defined as
\be
\chi^{00}({\bf q},\mathrm{i}\omega_n) = \frac{1}{2\beta}\int_{0}^{\beta} \! d\tau \,
\mathrm{e}^{\mathrm{i}\omega_n \tau}
\langle \mathcal{T}_{\tau} N_{{\bf q}}(\tau) N_{-{\bf q}}(0)\rangle 
\ee
with the density operator
\be
N_{{\bf q}}(\tau) = \frac{1}{\sqrt{\mathcal{N}}}\sum_{{\bf k},\mu,\sigma} c_{{\bf k} + {\bf q}\mu\sigma}^{\dagger}(\tau) c_{{\bf k}\mu\sigma}(\tau).
\ee
To derive RPA expressions for the above quantities in the absence of spin-rotation symmetry, it proves useful to introduce a generalized correlation function as
\be
\label{eq:genchi}
[\chi]^{\mu_1\sigma_1; \mu_2\sigma_2}_{\mu_3\sigma_3; \mu_4\sigma_4}({\bf q},{\bf q}^{\prime},\mathrm{i}\omega_n)=
\frac{1}{\beta\mathcal{N}}\int_{0}^{\beta} \! d\tau \, \mathrm{e}^{\mathrm{i}\omega_n \tau}\sum_{{\bf k},{\bf k}^{\prime}}
\langle \mathcal{T}_{\tau} 
c_{{\bf k} + {\bf q}\mu_1\sigma_1}^{\dagger}(\tau) 
c_{{\bf k}\mu_2\sigma_2}(\tau)
c_{{\bf k}^{\prime} - {\bf q}^{\prime}\mu_3\sigma_3}^{\dagger}(0) 
c_{{\bf k}^{\prime}\mu_4\sigma_4}(0)
\rangle,
\ee
since the transverse and longitudinal channels are in general coupled. We note that the correlation function \Eqref{eq:orbchi} in the main text is related to \Eqref{eq:genchi} as $[\chi]^{\mu_1\sigma_1; \mu_2\sigma_2}_{\mu_3\sigma_3; \mu_4\sigma_4}({\bf q},\mathrm{i}\omega_n) = \frac{1}{2}[\chi]^{\mu_1\sigma_1; \mu_2\sigma_2}_{\mu_3\sigma_3; \mu_4\sigma_4}({\bf q},{\bf q},\mathrm{i}\omega_n)$. To take into account Umklapp scattering processes in the computation of the RPA susceptibility, one introduces a matrix valued correlation function
\be 
[\chi_{l,l^{\prime}}]^{\mu_1\sigma_1 ;\mu_2\sigma_2}_{\mu_3\sigma_3;\mu_4\sigma_4}
({\bf q},\mathrm{i}\omega_n) \equiv
[\chi]^{\mu_1\sigma_1;\mu_2\sigma_2}_{\mu_3\sigma_3;\mu_4\sigma_4}
({\bf q} + {\bf Q}_{l},{\bf q} + {\bf Q}_{l^{\prime}},\mathrm{i}\omega_n).
\ee
The Umklapp indices $l$ and $l^{\prime}$ run over the values $0,1,2,3$ that correspond to the wavevectors ${\bf Q}_{0} = (0,0)$ and ${\bf Q}_{1} = (\pi,0)$, ${\bf Q}_{2} = (0,\pi)$, ${\bf Q}_{3} = (\pi,\pi)$. This structure is adapted to the case of double-Q order, but also contains the single-Q and paramagnetic cases.

The generalized correlation function can now be viewed as a matrix with a $4 \times 4$ block structure labeled by $l,l^{\prime}$, where each block is composed of a $100 \times 100$ matrix, where we put the orbital- and spin-index configurations $(\mu_1\sigma_1 ;\mu_2\sigma_2)$ and $(\mu_3\sigma_3;\mu_4\sigma_4)$ into a combined index.

The bare correlation function $[\chi_{l,l^{\prime}}^{0}]^{\mu_1\sigma_1;\mu_2\sigma_2}_{\mu_3\sigma_3;\mu_4\sigma_4}({\bf q},\mathrm{i}\omega_n) $ is computed as a bubble of Greens functions with the Hartree-Fock self-energy, $ G^{-1} =  [G^{0}]^{-1} - \Sigma_{\mathrm{HF}}$. The self-energy $\Sigma_{\mathrm{HF}}$ is specified by a $ 40 \times 40 $ matrix with respect to the $(l,\mu,\sigma)$ index tuple and carries all information about the magnetic state and the corresponding electronic band reconstruction. Performing the Matsubara sum one easily obtains
\be 
[\chi_{l,l^{\prime}}^{0}]^{\mu_1\sigma_1;\mu_2\sigma_2}_{\mu_3\sigma_3;\mu_4\sigma_4}({\bf q},\mathrm{i}\omega_n) & = &
 -\frac{1}{\beta\mathcal{N}}\int_{0}^{\beta} \! d\tau \,
\mathrm{e}^{\mathrm{i}\omega_n \tau}\sum_{\{l_{i}\}_{l,l^{\prime}}} \sum_{k}^{\prime}
 [G_{l_{2},l_{3}}]^{\mu_{2}\sigma_{2};\mu_{3}\sigma_{3}}({\bf k},\tau)
[G_{l_{1},l_{4}}]^{\mu_{1}\sigma_{1};\mu_{4}\sigma_{4}}({\bf k}-{\bf q},-\tau) \\
& = & -\frac{1}{\mathcal{N}}\sum_{{\bf k},n_1,n_2}^{\prime}
[\mathcal{M}_{n_1,n_2;l,l^{\prime}}({\bf k},{\bf q})]^{\mu_1\sigma_1;\mu_2\sigma_2}_{\mu_3\sigma_3;\mu_4\sigma_4}
\frac{f(\epsilon_{n_1}({\bf k}-{\bf q})) - f(\epsilon_{n_2}({\bf k}))}{\mathrm{i}\omega_n + \epsilon_{n_1}({\bf k}-{\bf q}) - \epsilon_{n_2}({\bf k})},
\ee
with the eigenenergies $\epsilon_{n}(\bf k)$ of the Hartree-Fock Hamiltonian and $f(\epsilon)$ the Fermi-Dirac distribution. The coherence factors entering the components of the bare correlation function read
\be 
[\mathcal{M}_{n_1,n_2;l,l^{\prime}}({\bf k},{\bf q})]^{\mu_1\sigma_1;\mu_2\sigma_2}_{\mu_3\sigma_3;\mu_4\sigma_4}=
\sum_{\{l_1,l_2,l_3,l_4\}_{l,l^{\prime}}}
\mathcal{U}_{{l_1\mu_1\sigma_1,n_1}}^{\ast}({\bf k} - {\bf q})
\mathcal{U}_{{l_2\mu_2\sigma_2,n_2}}({\bf k})
\mathcal{U}_{{l_3\mu_3\sigma_3,n_2}}^{\ast}({\bf k})
\mathcal{U}_{{l_4\mu_4\sigma_4,n_1}}({\bf k} - {\bf q}),
\ee
where $\sum_{\{l_1,l_2,l_3,l_4\}_{l,l^{\prime}}}$ denotes a restricted sum
over $l$-index tuples contributing to the $l$,$l^{\prime}$ component of the
correlation function and the prime on the sum denotes a ${\bf k}$-summation over the corresponding reduced Brillouin zone. The unitary matrix $\mathcal{U}_{{l\mu\sigma,n}}({\bf k})$ diagonalizes the Hartree-Fock Hamiltonian. The spinor $\Psi_{{\bf k}l\mu\sigma}$ (where now we have made all quantum numbers explicit) transforms as $\Psi_{{\bf k}l\mu\sigma} = \sum_{n}\mathcal{U}_{{l\mu\sigma,n}}({\bf k}) \Phi_{{\bf k}n}$. Without spin-rotational symmetry, the spin index $\sigma$ is in general not a conserved quantum number. If spin is still approximately conserved, the single-particle eigenstates of the Hartree-Fock Hamiltonian will still have a large overlap with exact spin eigenstates. In this case, the additional components of the generalized correlation function with $\sigma_{1} \neq \sigma_{4}$ and $\sigma_{2} \neq \sigma_{3}$ are expected to be small. If the system has $\mathrm{SU}(2)$ or $\mathrm{U}(1)$ spin-symmetry, only the $\sigma_{1} = \sigma_{4}$, $\sigma_{2} = \sigma_{3}$ components of the bare generalized correlation function are non-zero.

The RPA equation for the generalized correlation function can be derived from, e.g., the equation of motion approach and an appropriate truncation of the resulting hierarchy of equations. From this, we obtain the determining equation as
\be 
\label{eq:RPAequation}
[\chi_{l,l^{\prime}}]^{\mu_1\sigma_{1};\mu_2\sigma_{2}}_{\mu_3\sigma_{3}; \mu_4\sigma_{4}}({\bf q},\mathrm{i}\omega_n) =
[\chi_{l,l^{\prime}}^{0}]^{\mu_1\sigma_{1};\mu_2\sigma_{2}}_{\mu_3\sigma_{3};\mu_4\sigma_{4}}({\bf q},\mathrm{i}\omega_n) +
[\chi_{l,m}^{0}]^{\mu_1\sigma_{1};\mu_2\sigma_{2}}_{\nu_1\tau_{1};\nu_2\tau_{2}}({\bf q},\mathrm{i}\omega_n)
[U]^{\nu_1\tau_{1};\nu_2\tau_{2}}_{\nu_3\tau_{3};\nu_4\tau_{4}}
[\chi_{m,l^{\prime}}]^{\nu_3\tau_{3};\nu_4\tau_{4}}_{\mu_3\sigma_{3};\mu_4\sigma_{4}}({\bf q},\mathrm{i}\omega_n).
\ee
Repeated indices are summed over in \Eqref{eq:RPAequation}. The bare fluctuation vertex $[U]^{\nu_1\tau_{1};\nu_2\tau_{2}}_{\nu_3\tau_{3};\nu_4\tau_{4}}$ originates from the Hubbard-Hund interaction \Eqref{eq:interaction}, and describes how electrons scatter off a collective excitation in the particle-hole channel. Since we employ the Hubbard-Hund interaction with interaction parameters preserving spin-rotational symmetry, it is still possible to classify the scattering of collective excitations according to their total spin. Accordingly, the vertex can be split into three different contributions as
\be 
[U]^{\nu_1\tau_{1};\nu_2\tau_{2}}_{\nu_3\tau_{3};\nu_4\tau_{4}} =
[U_{1}]^{\nu_1\tau_{1};\nu_2\tau_{2}}_{\nu_3\tau_{3};\nu_4\tau_{4}} +
[U_{2}]^{\nu_1\tau_{1};\nu_2\tau_{2}}_{\nu_3\tau_{3};\nu_4\tau_{4}} +
[U_{3}]^{\nu_1\tau_{1};\nu_2\tau_{2}}_{\nu_3\tau_{3};\nu_4\tau_{4}},
\ee
where $U_{1}$ and $U_{3}$ describe the scattering of opposite spin and equal spin fluctuations in the longitudinal channel, respectively, while $U_{2}$ describes the scattering of transverse spin fluctuations. But since the RPA equation couples the different excitation channels, longitudinal and transverse excitations will in general be mixed.

The vertex contribution $U_{1}$ is defined as
\be
[U_{1}]^{\mu\sigma;\mu\sigma}_{\mu\bar{\sigma};\mu\bar{\sigma}} = - U,
\quad
[U_{1}]^{\mu\sigma;\mu\sigma}_{\nu\bar{\sigma};\nu\bar{\sigma}} = - U^{\prime},
\quad
[U_{1}]^{\mu\sigma;\nu\sigma}_{\nu\bar{\sigma};\mu\bar{\sigma}} = - J,
\quad
[U_{1}]^{\mu\sigma;\nu\sigma}_{\mu\bar{\sigma};\nu\bar{\sigma}} = - J^{\prime},
\quad\text{with}\,\mu \neq \nu 
\ee
where $\bar{\sigma}$ denotes the opposite spin polarization to $\sigma$.  The $U_{1}$ contribution is zero for all other orbital or spin index combinations. For the equal spin fluctuation vertex, we find the non-zero elements
\be
[U_{3}]^{\mu\sigma;\mu\sigma}_{\nu\sigma;\nu\sigma} =  -(U^{\prime} - J),
\quad
[U_{3}]^{\nu\sigma;\mu\sigma}_{\mu\sigma;\nu\sigma} = (U^{\prime}-J),
\quad\text{with}\,\mu \neq \nu.
\ee
For the transverse channel, we obtain
\be
[U_{2}]^{\mu\bar{\sigma};\mu\sigma}_{\mu\sigma;\mu\bar{\sigma}} = U,
\quad
[U_{2}]^{\nu\bar{\sigma};\mu\sigma}_{\mu\sigma;\nu\bar{\sigma}} = U^{\prime},
\quad
[U_{2}]^{\nu\bar{\sigma};\nu\sigma}_{\mu\sigma;\mu\bar{\sigma}} = J,
\quad
[U_{2}]^{\mu\bar{\sigma};\nu\sigma}_{\mu\sigma;\nu\bar{\sigma}} = J^{\prime},
\quad\text{with}\,\mu \neq \nu,
\ee
and zero else. For (residual) continuous spin-rotational symmetry, the transverse and longitudinal channels decouple and can be treated independently.

The linear matrix equation \Eqref{eq:RPAequation} is then solved numerically by matrix inversion. From the $ l = l^{\prime} = 0 $ component of the generalized correlation function $[\chi_{l,l^{\prime}}]^{\mu_1\sigma_1;\mu_2\sigma_2}_{\mu_3\sigma_3;\mu_4\sigma_4}
({\bf q},\mathrm{i}\omega_n) $ the spin susceptibilities $ \chi^{ij}({\bf q},\mathrm{i}\omega_n) $ and the density susceptibility $ \chi^{00}({\bf q},\mathrm{i}\omega_n) $ can be recovered by forming the appropriate linear combinations. For the sake of completeness, we also give our definition of the transverse susceptibility 
\be
\chi^{+-}({\bf q},\mathrm{i}\omega_n) &  = &  
\frac{1}{2}\sum_{\mu,\nu}[\chi_{0,0}]^{\mu\uparrow;\mu\downarrow}_{\nu\downarrow;\nu\uparrow}({\bf q},\mathrm{i}\omega_n) \\ 
& = &
\frac{1}{2\beta\mathcal{N}}\sum_{\mu,\nu}\int_{0}^{\beta} \! d\tau \, \mathrm{e}^{\mathrm{i}\omega_n \tau}\sum_{{\bf k},{\bf k}^{\prime}}\langle \mathcal{T}_{\tau} 
c_{{\bf k} + {\bf q}\mu\uparrow}^{\dagger}(\tau) 
c_{{\bf k}\mu\downarrow}(\tau)
c_{{\bf k} - {\bf q}^{\prime}\nu\downarrow}^{\dagger}(0) 
c_{{\bf k}\nu\uparrow}(0)
\rangle.
\ee
The spectral density is obtained from performing the analytic continuation $\mathrm{i}\omega_{n} \to \omega + \mathrm{i} \eta$ with some finite but small smearing parameter $ \eta > 0 $ on the order of the temperature $k_{\mathrm{B}}T = \beta^{-1}$ and taking the imaginary part.

While the physical susceptibilities are composed only of the intraorbital contributions $\mu_{1} = \mu_{2}$ and $\mu_{3} = \mu_{4}$, we note that the interorbital components also contain information about interorbital order-parameter fluctuations. We therefore also introduce the inverse fluctuation propagator
\be
[\mathcal{D}_{l,l^{\prime}}^{-1}]^{\mu_1\sigma_{1};\mu_2\sigma_{2}}_{\mu_3\sigma_{3}; \mu_4\sigma_{4}}({\bf q},\mathrm{i}\omega_n) = 
[U^{-1}]^{\mu_1\sigma_{1};\mu_2\sigma_{2}}_{\nu_1\tau_{1};\nu_2\tau_{2}} 
\left[\delta_{l,l^{\prime}}\delta^{\nu_{1}\mu_{3}}\delta^{\nu_{2}\mu_{4}}\delta^{\tau_{1}\sigma_{3}}\delta^{\tau_{2}\sigma_{4}} +  [U]^{\nu_1\tau_{1};\nu_2\tau_{2}}_{\nu_3\tau_{3};\nu_4\tau_{4}}
[\chi_{l,l^{\prime}}]^{\nu_3\tau_{3};\nu_4\tau_{4}}_{\mu_3\sigma_{3}; \mu_4\sigma_{4}}({\bf q},\mathrm{i}\omega_n) 
\right]
\ee 
describing the propagation of a collective excitation in the particle-hole channel with arbitrary orbital character.

\section{Crossover scales between orbitally uniform and orbitally polarized regimes}
\label{app:orbpol}

Here we briefly describe how the crossover lines in Fig.~\ref{fig:RPA_panel}(f), (l), (r) are obtained from our data. The extracted crossover scales separate the orbitally uniform regime from the orbitally polarized regime. We find these energy scales to be different between the $C_{2}$ MS state and the $C_{4}$ SCO and OM states. In Fig.~\ref{fig:supppanel}(a),(f),(k) we show the total spectral weight of magnetic excitations for MS, SCO and OM, respectively.
\begin{figure*}[th]
\includegraphics[width=1\columnwidth]{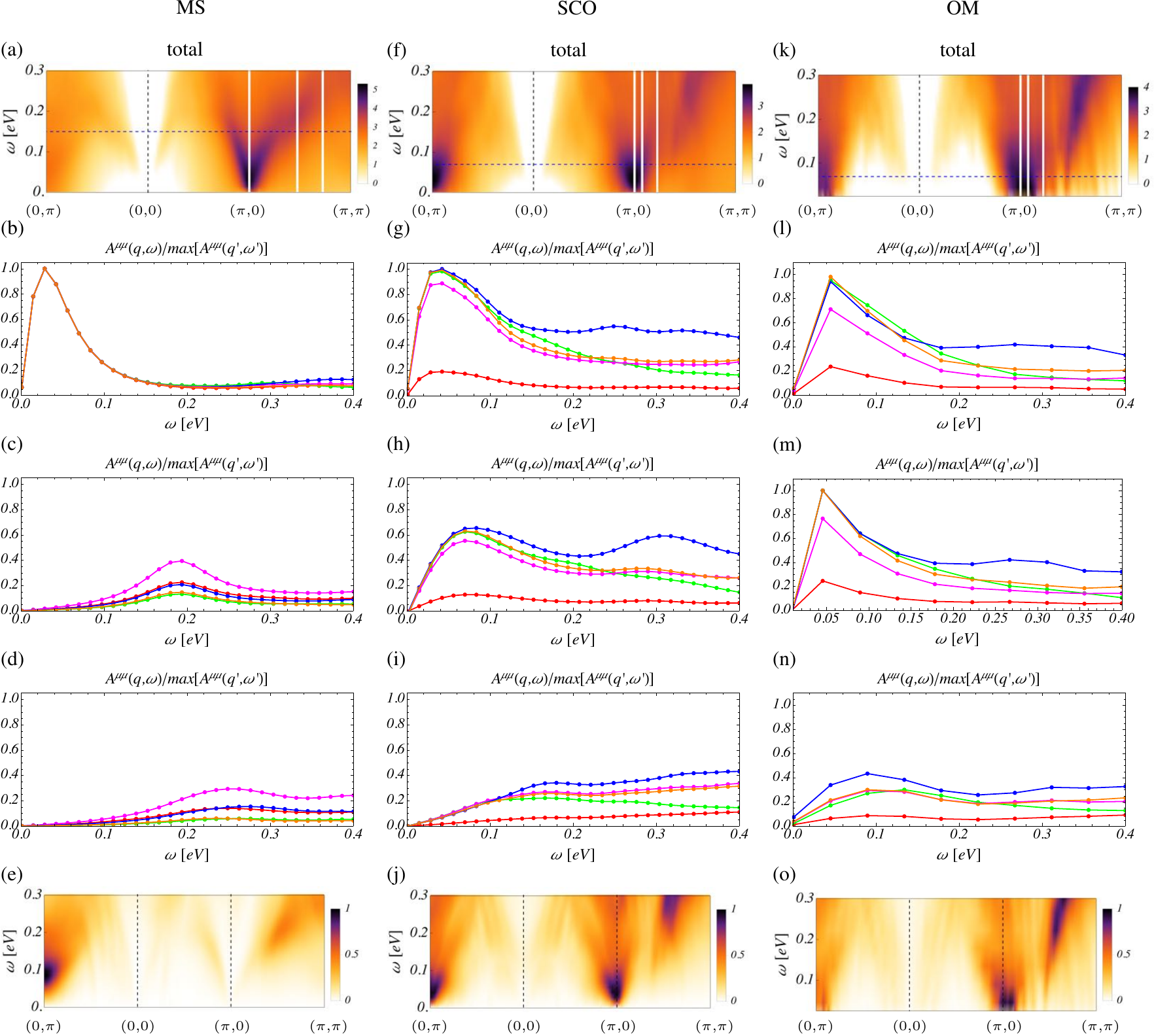}
\caption{(a), (f), (k) Cut through the total spectral weight distribution $\mathcal{A}({\bf q},\omega)$ of magnetic excitations in units of $1/\mathrm{eV}$ along the high symmetry momentum-space path $(\pi,0)-(0,0)-(0,\pi)-(\pi,\pi)$. The colorscale corresponds to the log of $\mathcal{A}({\bf q},\omega)$. The blue dashed vertical lines mark the crossover between orbitally uniform and orbitally polarized regime of the gapless excitation branch on the energy axis (a) $150 \, \mathrm{meV}$ (f) $ 70 \, \mathrm{meV} $, (k) $ 70 \, \mathrm{meV} $ (note the offset from $\omega = 0 $ in (k) on the vertical axis). The white lines mark the cuts, along which the (normalized) orbitally resolved spectral weight is shown in (b) - (d) for the MS state, in (g) - (i) for the SCO state and (l) - (n) for the OM state. The colors of the curves encode the orbital composition of the magnetic excitations as $d_{xz}$ ({\color{red} red}) ,  $d_{yz}$ ({\color{green} green}), $d_{x^{2}-y^{2}}$ ({\color{magenta} magenta}), $d_{xy}$ ({\color{blue} blue}), $d_{3z^{2}-r^{2}}$ ({\color{orange} orange}). The curves are normalized to the maximum of the respective orbital component for all ${\bf q}$ and $\omega$ in the momentum-frequency grid. (e), (j), (o) The quantity $\Delta({\bf q},\omega)$ as a measure of orbital polarization.}
\label{fig:supppanel}
\end{figure*}
The white lines mark cuts along the energy axis for fixed momenta along which we plot the (normalized) orbitally resolved spectral weight in Fig.~\ref{fig:supppanel}(b)-(d) for the MS state, (g)-(i) for the SCO state and (l)-(n) for the OM state. The first plot in each case corresponds to the cut along $(\pi,0)$. The second plot corresponds to the middle cut, while the third plot correspond to the rightmost cut. The normalization is with respect to $\mathrm{max}_{{\bf q},\omega} \mathcal{A}^{\mu\mu}({\bf q},\omega)$ with ${\bf q}$ and $\omega$ spanning the cut through the excitation spectrum. We define the normalized spectral weight
\be
w^{\mu}({\bf q},\omega) = \mathcal{A}^{\mu\mu}({\bf q},\omega)/\mathrm{max}_{{\bf q}^{\prime},\omega^{\prime}} \mathcal{A}^{\mu\mu}({\bf q}^{\prime},\omega^{\prime}).
\ee
The maxima in $w^{\mu}({\bf q},\omega)$ roughly track the dispersion of the spinwave excitations. For the MS state, the low-energy excitations at $(\pi,0)$ feature no orbital polarization. Only for higher energies above an energy of $\sim 150 \, \mathrm{meV}$ do we find a splitting of the orbitally resolved spectral weight. As we probe the incoherent excitations, see Fig.~\ref{fig:supppanel}(d), the orbital polarization becomes more pronounced over a wider energy window. The normalized spectral weight for SCO and OM features a splitting of the curves already for low energies. As the momentum moves away from the ordering vector, the degree of orbital polarization seems to decrease.

As a measure of orbital polarization, we introduce the quantity 
\be
\Delta({\bf q},\omega) = \mathrm{max}_{\mu} w^{\mu}({\bf q},\omega) - \mathrm{min}_{\mu} w^{\mu}({\bf q},\omega)
\ee
that we plot in Fig.~\ref{fig:supppanel}(e), (j), (o) for MS, SCO and OM states, respectively. In the MS state, only the gapped mode at $(0,\pi)$ and the high-energy branches around $(\pi,0)$ feature an appreciable orbital polarization. For SCO and OM, the situation is reversed: at the ordering vectors, the gapless modes feature strong orbital polarization. As the momentum deviates from the ordering vectors, the orbital polarization decreases rapidly. On the energy axis, the orbital polarization persists, but decreases around  
$ 70 \, \mathrm{meV}$.



\end{widetext}

\end{document}